\documentclass[english,aps,preprint,showpacs,amsmath,amssymb]{revtex4}
\usepackage{graphicx,dcolumn,bm,bbm,amssymb,babel,multirow,times}
\newcommand {\oops}[1]{%
\textbf{#1}}
\begin{document}

\begin{flushright}
{\it Nature Communications 2, 170 (2011)}
\end{flushright}

\vspace{0.5cm}
{\Large
\noindent
\oops{Rescuing ecosystems from extinction cascades through compensatory perturbations}\\
}

\oops{Sagar Sahasrabudhe$^1$ \& Adilson E. Motter$^{1,2}$}\\

\noindent
{\it $^1$ Department of Physics and Astronomy, Northwestern University, Evanston, IL 60208, USA}

\noindent
{\it $^2$ Northwestern Institute on Complex Systems, Northwestern University, Evanston, IL 60208, USA}\\

{\bf
Food-web perturbations stemming from climate change, overexploitation, invasive species, and habitat degradation often cause an initial loss of species that results in a cascade of secondary extinctions, posing considerable challenges to ecosystem conservation efforts. Here we devise a systematic network-based approach to reduce the number of secondary extinctions using a predictive modeling framework.
We show that the extinction of one species can often be compensated by the concurrent {\it removal} or {\it population suppression} of other specific species, which is a counterintuitive effect not previously tested in complex food webs. These compensatory perturbations frequently involve long-range interactions that are not evident from local predator-prey relationships. In numerous cases, even the early removal of a species that would eventually be extinct by the cascade is found to significantly reduce the number of cascading extinctions. These compensatory perturbations only exploit resources available in the system, and illustrate the potential of human intervention combined with predictive modeling for ecosystem management.\\
}

Halting the loss of biodiversity caused by human and natural forces \cite{thomas04,Je,Pi,landmammals} has become one of the grand challenges of this century. Despite the evolutionarily acquired robustness of ecological systems, the disappearance or significant suppression of one or more species can propagate through the food-web network and cause other species to go extinct as the system approaches a new stable state \cite{Pa,Sc}. A well-documented example is the trophic cascade observed over the past 40 years in the coastal northwestern Atlantic Ocean, where the depletion of great sharks released cownose ray, whose enhanced predation on scallop has driven the latter to functional extinction in some areas \cite{My}. The massive extinction of terrestrial and freshwater species, including butterflies, birds, fishes and mammals, that started in Singapore in the early 1800s is a striking example of an extinction cascade caused by heavy deforestation \cite{Ba}. Invasive species, such as e
 xotic aquatic species introduced by ballast water transported in commercial ships, are yet another frequent cause of extinction of native species \cite{Cy}. These species alter the food-web structure and dynamics \cite{Ro}, leading to potentially devastating long-term effects for the local ecosystem \cite{Pi}.

A number of studies have been conducted on the prediction and analysis of secondary extinctions \cite{Ebenman}, both structural \cite{Stefano,Jennifer,Richard,Sole,Coll} and dynamical \cite{Anna,Owen,Quince,Borrvall,Ives,berlow2009}, after the loss of one species. However, there is a fundamental lack of understanding on how these cascades of secondary extinctions could be mitigated. Different approaches to prevent species extinctions have been proposed in previous studies \cite{Ch}, including the eradication or seasonal removal of a predator of a species that is in danger of extinction and, in few cases, the control of a population that is not in direct interaction with the species meant to be protected  \cite{Wo}. However, in most of these efforts the aim has been to save {\it one} species---generally a visibly endangered one \cite{redlist}---at the potential expense of others.  These interventions do not usually account for cascading effects and at times have been found to 
 have an impact that was the opposite of the desired one \cite{rayner07}.  Due to the integrated nature of food-web systems, a species that does not exhibit a feeding interaction with some other species can still have substantial influence on the other species' population.  Yet, the possibility of exploiting this inherent complexity to prevent multiple extinctions has not yet been pursued.  

Here, inspired by recent advances in the control of complex physical and biochemical networks \cite{motter04,motter08}, we study mechanisms by which extinction cascades can be mitigated and identify compensatory perturbations that can rescue otherwise threatened species {\it downstream} the cascades. These compensatory perturbations consist of the concurrent removal, mortality increase or growth suppression of target species, which, as discussed below, are interventions that have a strong empirical basis and can in principle prevent most or all secondary extinctions.\\  

\noindent
{\large\bf Results}

\begin{figure}[tbh] 
\includegraphics[scale=0.27]{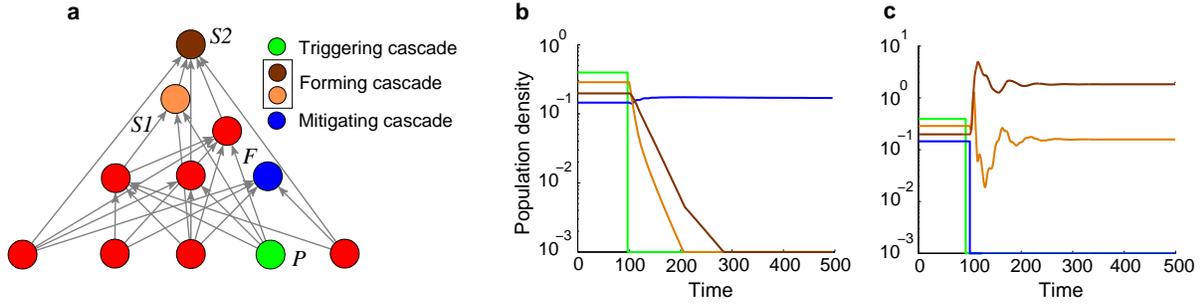}
\caption{ \baselineskip 14pt
\footnotesize{{\bf Example of the impact of species removal.}
({\bf a}) The removal of a basal species, $P$, triggers a cascade that leads to the subsequent extinction of two high-trophic species, $S1$ and $S2$, in this initially persistent $11$-species food web. The removal of an intermediate-trophic  species, $F$, shortly after the removal of $P$ prevents the propagation of the cascade,
and causes no additional extinctions. ({\bf b}, {\bf c}) The time evolution of the populations following the removal of $P$ ({\bf b}) and following the combined removal of $P$ and  $F$ ({\bf c}) shows that the otherwise vanishing populations of $S1$ and $S2$ can reach stationary levels comparable to or higher than the unperturbed ones in a time-scale of the order of the time-scale of the cascade (color code defined in ({\bf a})). The long-range character of the underlying interactions is emphasized by the fact that species $F$ is not directly connected to either the species triggering the cascade or the ones rescued. This food web was simulated using the consumer-resource model (Supplementary Methods).
}
\label{fig1}}
\end{figure}

\noindent
{\bf Rescue mechanism.}
The proposed rescue mechanism is illustrated in Figure~\ref{fig1}.  In this example, the sudden extinction of species $P$ leads to the subsequent extinction of species $S1$ and $S2$ (Fig.~\ref{fig1}b).  However, the proactive removal of species $F$ shortly after the initial extinction drives the system to a new stable state in which no additional species are extinct (Fig.~\ref{fig1}c).  The initial extinction, which we refer to as the \underline{p}rimary removal ($P$), models the initial perturbation, whereas the proactive removal is the compensatory perturbation that we seek to identify. We refer to the latter as the \underline{f}orced removal ($F$). In this case, it prevents all secondary extinctions and leads to a system with 10 instead of 9 persistent species.  The absence of feeding interactions between species $F$ and the species involved in the cascade (Fig.~\ref{fig1}a)  illustrates the limitations of conclusions derived from direct inspection of the food-web structur
 e, and emphasizes the importance of a modeling framework that can account for both the nonlinear and the system-level nature of the network response to perturbations. Following this example, we first consider the rescue effects of total species removals. Below we relax this condition to also consider partial removals and other interventions.

To explore the principle underlying the example of Figure~\ref{fig1}, we developed an algorithm that we use to systematically identify compensatory perturbations. This is implemented using two well-established models to describe the dynamics: the multi-species consumer-resource model \cite{Yo,Mar}, which allows for adaptive behavior of the predators and takes into account different types of functional responses; and the predator-prey Lotka-Volterra model, which assumes a linear approximation for the interaction coefficients and does not involve adaptive strategies \cite{Zh} (see details in Supplementary Methods).
While the former is potentially more realistic, the Lotka-Volterra model allows for more thorough analysis. Our algorithm is based on identifying the fixed points  $(X_1^*, \dots ,X_n^*)$ of the post-perturbation dynamics, which in the Lotka-Volterra case are given by  $X_i(b_i+\sum_{j}a_{ij}X_j)=0$, where $X_i\ge 0$ and $b_i$ represent respectively the population and mortality rate (or growth rate, in the case of the first trophic level) of species $i$, and $a_{ij}$ represents the food-web structure. These fixed points are time-independent solutions that we use as target states to design compensatory perturbations whenever the number of extinct species is reduced at one such point.
Specifically, we proceed as follows: ({\it i}) we start with a primary species removal on an initially persistent food web that, according to our model dynamics, is predicted to lead to secondary extinctions; ({\it ii}) we identify the fixed points of the dynamics under the constraint imposed by the primary removal---these fixed points typically have one or more species with zero population, in addition to the one corresponding to the primary removal; ({\it iii}) starting from fixed points with the largest number of positive populations, we test the impact of the forced removal of a species that has zero population at the fixed point. This last step is implemented immediately after the primary removal and is repeated over different fixed points until the most effective forced removals are identified. For more details on the rescue algorithm see Methods.

A similar algorithm is implemented in the case of the consumer-resource model except that, because the characterization of the asymptotic dynamics is in that case more involved, the identification of the rescues is done by exhaustive search over all possible forced removals and subsequent selection of the removals that minimize the number of extinctions. 
In either case, the forced removals are tailored  to drive the system to a fixed point if the point is stable, or to the corresponding neighborhood if the point is unstable (Supplementary Methods; Supplementary Figs.\ S1-S3).
We have implemented the proposed approach using both model and empirically-observed food webs.  

\medskip
\noindent
{\bf Model food webs.} 
Our model food webs were generated using the niche model \cite{Wi2000}, which is based on ecologically relevant principles
(Supplementary Methods).
Within this model, we considered extinction cascades triggered by the primary removal of one species in initially persistent food webs \cite{Stouffer}. As a compromise between computational feasibility and complexity, we focused mainly on food webs of 15 species generated from a connectance of 0.2, which, in the case of the consume-resource dynamics, were taken to have a mixed vertebrate-invertebrate community type (Supplementary Methods).
The evolution of these food webs can be either time-dependent or -independent and will generally depend on the structure of the network, parameter choice, and dynamical model \cite{Williams2009}. We also analyzed the impact of systematically varying the food-web parameters and the size of the primary perturbation (see below). In all cases, the number of cascades mitigated by the forced removal of one species is found to be comparable or larger than the number of cascades not mitigated 
(Supplementary Methods),
which provides evidence that the proposed procedure applies to diverse systems. For detailed statistics on the number of rescued species, see Supplementary Methods and Supplementary Figure S4.
But what are the network mechanisms affording these rescue interactions?

\begin{figure}[h]
\includegraphics[scale=0.45]{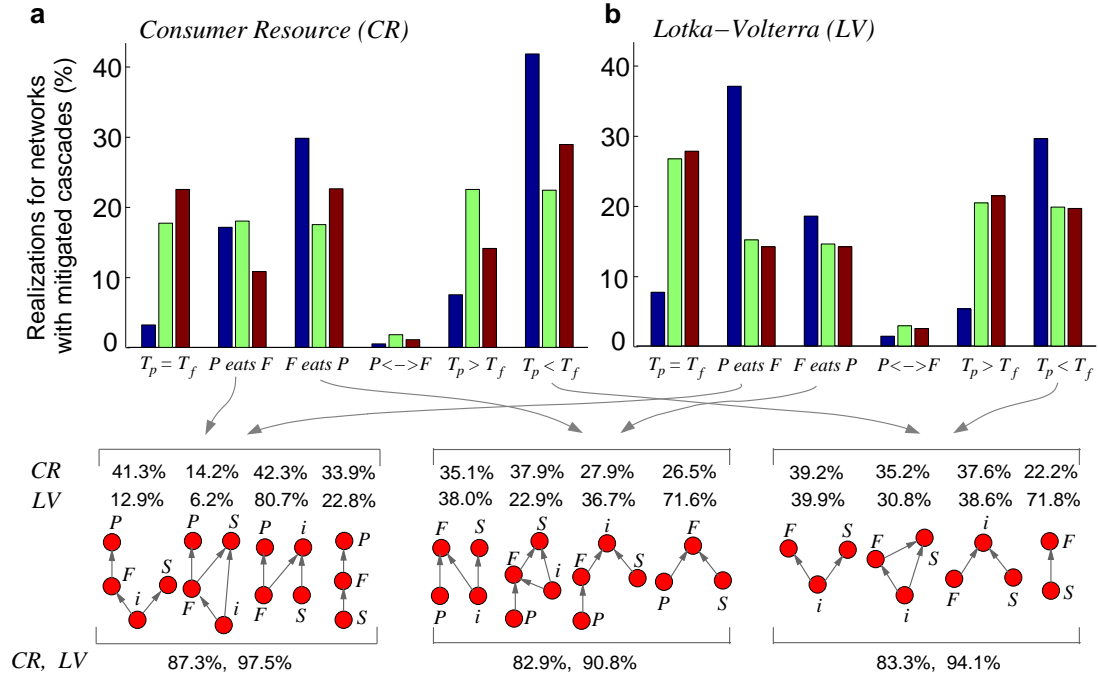}
\caption{ \baselineskip 14pt
\footnotesize{{\bf Rescue interactions predicted for model food webs.}
({\bf a}, {\bf b}) Classification of the mitigated cascades (blue) 
according to the feeding relations between the primary removal $P$ 
and forced removal $F$, and associated baseline models where $F$ (brown) 
or both $P$ and $F$ (green) are replaced  by random species.
The statistics are for forced removals that rescue the largest number of species in networks of $15$ species. 
The relations are organized according to whether $P$ and $F$ are in the same trophic level ($T_p=T_f$), share a predator-prey link ($P$ eats $F$ or $F$ eats $P$), have a reciprocal predator-prey interaction ($P\! <\!\!\!-\!\!\!>\! F$), or are in different trophic levels without being directly connected ($T_p>T_f$ or $T_p<T_f$). For simplicity, we use the `predator-prey' terminology also when the feeding interactions involve basal species. The bottom sets indicate the local network structure of the scenarios that have a frequency higher than one or both of the baselines, where $S$ indicates a species that would be part of the extinction cascade and is rescued by the removal of $F$: if $F$ and $S$ share a common prey $i$, then the removal of $F$ causes an increase in $i$ (in $98.0\%$ ({\bf a}) and $98.3\%$ ({\bf b}) of the occurrences if $F$ and $S$ are not connected and $82.5\%$ ({\bf a}) and $98.9\%$ ({\bf b}) if $S$ feeds on $F$ as well), which helps sustain $S$; if $F$ and $S$ share a common predator $i$, then the removal of $F$ causes a decrease in $i$ (in $57.1\%$ ({\bf a}) and
  $87.6\%$ ({\bf b}) of the occurrences), which often reduces predation of $S$; a fourth case is when $F$ is directly feeding on $S$, and hence its removal tends to enhance $S$ ($F$ would otherwise increase following the removal of $P$ in $54.0\%$ ({\bf a}) and $70.1\%$  ({\bf b}) of such occurrences).  
The percentages in the bottom sets indicate the fractions of forced removals in the corresponding category that take part in the network structures shown. The total percentages (bottommost) also
account for co-occurrences of different such structures for the same forced removal.
}
\label{fig2}} 
\end{figure}

Figure~\ref{fig2} shows the feeding relations between the primary removal $P$ and forced removal $F$ in the model food webs whose cascades are mitigated. In this figure, as well as in other parts of this study, the trophic levels are estimated using the prey-averaged trophic level algorithm \cite{williams04}. By comparing these two species with baselines in which one or both are replaced by randomly selected species,  we demonstrate that rescue interactions are more likely than at least one of the baselines  for (I) $P$ {\it feeding on} $F$, (II) $F$ {\it feeding on} $P$, and (III) $F$ {\it at a higher trophic level than} $P$ (while not sharing a predator-prey link with $P$).  These are also the most common scenarios in absolute numbers, accounting for more than $85\%$ of all rescues for both the consume-resource (Fig.~\ref{fig2}a) and the Lotka-Volterra (Fig.~\ref{fig2}b) dynamics. The two dynamical models also exhibit significant differences. The most fundamental difference
 , which follows from a direct comparison between the baseline models, is that $P$ is biased towards lower trophic levels for the consumer-resource model while it is distributed more uniformly for the Lotka-Volterra model. This explains the larger frequency of rescues for scenario (III) in Figure~\ref{fig2}a when compared to Figure~\ref{fig2}b, and at least part of the difference for scenarios (II) and (I). The observed difference in the distribution of $P$ is most likely due to the adaptive strategy inherent to the consumer-resource dynamics. (Another difference evident from Figure~\ref{fig2} is that the Lotka-Volterra case exhibits a larger number of rescues for $P$ and $F$ in the same trophic level, but this is mainly because the (initially persistent) Lotka-Volterra networks tend to have a larger number of basal species than the consumer-resource networks (Supplementary Methods)).

The higher than-by-chance frequency of scenario (I) indicates that cascades can often be mitigated by suppressing a low-trophic species released by the initial perturbation. Surprisingly, examination of the local network structure reveals that this suppression is more frequently mediated by a predator or a prey that is common to both $F$ and a rescued species $S$ than through a direct predator-prey link between $F$ and $S$ (Fig.~\ref{fig2}, bottom sets). In some cases the released species is a mesopredator \cite{crooks99}, but we note that for scenario (I) species $F$ is frequently basal ($26\%$ of the cases in Figure~\ref{fig2}a and $75\%$ of the cases in Figure~\ref{fig2}b, when averaged over the potentially non-unique $F$ that reduce the most a cascade triggered by a given $P$). For $P$ feeding on $F$, over $87\%$ of the $P$-$F$ pairs in
Figure~\ref{fig2}a and $97\%$ in Figure~\ref{fig2}b involve at least one network structure in which $F$ is related to $S$ either directly or by a common predator or prey. Similar structures and statistics are found for $P$ and $F$ exhibiting different feeding relations. This holds, in particular, for scenario (III), where the interaction between $P$ and $F$ is also indirect.

To further clarify their role, in Figure~\ref{fig2} we also analyze the {\it dynamics} on these local network structures shortly after the removal perturbations. For example, more than $98\%$ of all cases shown in which $F$ and $S$ share a common prey $i$ and $S$ does not feed on $F$, the removal of $F$ increases the population of $i$, which tends to favor $S$. 
(Such percentages were calculated one time unit after the perturbations and were found to be highly correlated with the corresponding asymptotic behavior.)
More interesting, in over $57\%$ of the cases in Figure~\ref{fig2}a and $87\%$ in Figure~\ref{fig2}b where $F$ and $S$ share a common {\it predator} $i$, the removal of $F$ rescues $S$ while decreasing the population of $i$, which indicates that the loss of a prey tends not to be fully transferred to the remaining preys even in the (adaptive) consumer-resource model. This transfer effect has been found to be relevant in specific case studies, such as in the fox-pig-eagle food web of the Channel Islands \cite{collins2009}. These mechanisms are not exhaustive and other long-range interactions are likely to play a role, as illustrated by the fact that over $30\%$ of the mitigated cascades in scenario (III) involve a situation in which a species is rescued upon removal of one of its preys; a related, longer-range ``$S$ feeding on $F$" structure is identified below in the case of the Chesapeake Bay food web.

\begin{figure}[t]
\includegraphics[scale=0.28]{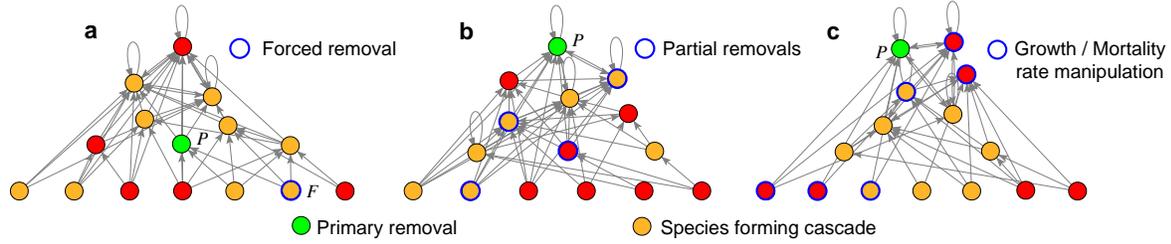}
\caption{ \baselineskip 14pt
\footnotesize {{\bf Examples of rescue interactions generated by
optimally compensatory interventions.} ({\bf a}) The removal of one species (green) leads to the extinction of $9$ other species (yellow). The concurrent forced removal of one of the otherwise vanishing species (blue ring) prevents all other cascading extinctions.  ({\bf b}) The removal of one species causes the extinction of $7$ other species. The partial removal of $4$ species (blue rings) prevents all cascading extinctions. ({\bf c}) The removal of one species leads to the extinction of $8$ other species. The permanent reduction of the growth rate of $3$ basal species and increase of the mortality rate of $3$ non-basal species (blue rings) prevent all cascading extinctions. These food webs were simulated using the Lotka-Volterra dynamics. Note that in all cases there is an overlap between the species forming the cascade and the ones that are proactively manipulated to prevent the cascade.}
\label{fig3}}
\end{figure}

\medskip
\noindent
{\bf Nondestructive interventions.}
Having shown that the locally deleterious removal of a species can have a net positive global impact in the imminence of an extinction cascade, we now consider three strictly nondestructive interventions. First, under appropriate conditions, the early removal of a species that would otherwise be eventually {\it extinct} by the cascade can prevent {\it all other} secondary extinctions (Methods). One such example is given in Figure~\ref{fig3}a, where the primary removal of species $P$ causes the subsequent extinction of $9$  species, but the removal of (the cascading) species $F$ shortly after the initial perturbation drives the system to a stable fixed point where all other populations are positive. This is a dramatic example of how the fate of a food web can depend on the order and timing of the events as much as it does on the events themselves. 

Second, the {\it partial removal} of one or more species can often prevent {\it all} secondary extinctions. This is generally possible if after the primary removal the dynamics has a fixed point in which a) all other populations are positive and b) the populations of one or more species are smaller than the corresponding populations at the early post-perturbation state.
The rescue intervention then consists on partial removals of these species to reduce their populations to those of the target fixed point (Methods).
This case is exemplified in Figure~\ref{fig3}b, where the partial removal of $4$ species fully compensates for the perturbation caused by the primary removal of species $P$, and rescues all $7$ otherwise vanishing species. 

Third, the {\it manipulation of the growth and mortality rates} of basal and non-basal species, respectively, is another intervention that can prevent {\it all} secondary extinctions. We assume that growth rates can often only be decreased and mortality rates can only be increased, which, like in the case of species removal, can be achieved by only exploiting 
natural resources available in the system.  We consider all such changes that lead to time-independent dynamics (zero time derivative) for the populations of the corresponding species shortly after the initial perturbation (Methods).
These interventions are designed to reduce the likelihood that these populations will oscillate or decrease to zero. This case is illustrated in Figure~\ref{fig3}c, where the secondary extinctions of $8$ species triggered by the removal of $P$ are prevented by manipulating the growth/mortality rates of $6$ species. The statistics for total and partial cascade prevention are summarized in Supplementary Table S1 and Supplementary Figure S4.

\medskip
\noindent
{\bf Empirically-observed food webs.}
The empirically-observed networks we considered are the Chesapeake Bay food web \cite{Ul}, an aquatic network with 33 
species, and the Coachella Valley food web \cite{Po}, a terrestrial network with 30 species, both modeled using the consumer-resource dynamics (Supplementary Methods).
These systems, as many other empirically reconstructed food webs, are relatively robust against perturbations. To generate an appreciable number of cascades, these networks were perturbed by the primary removal of three rather than one species. 

\begin{figure}[h]
\includegraphics[scale=0.34]{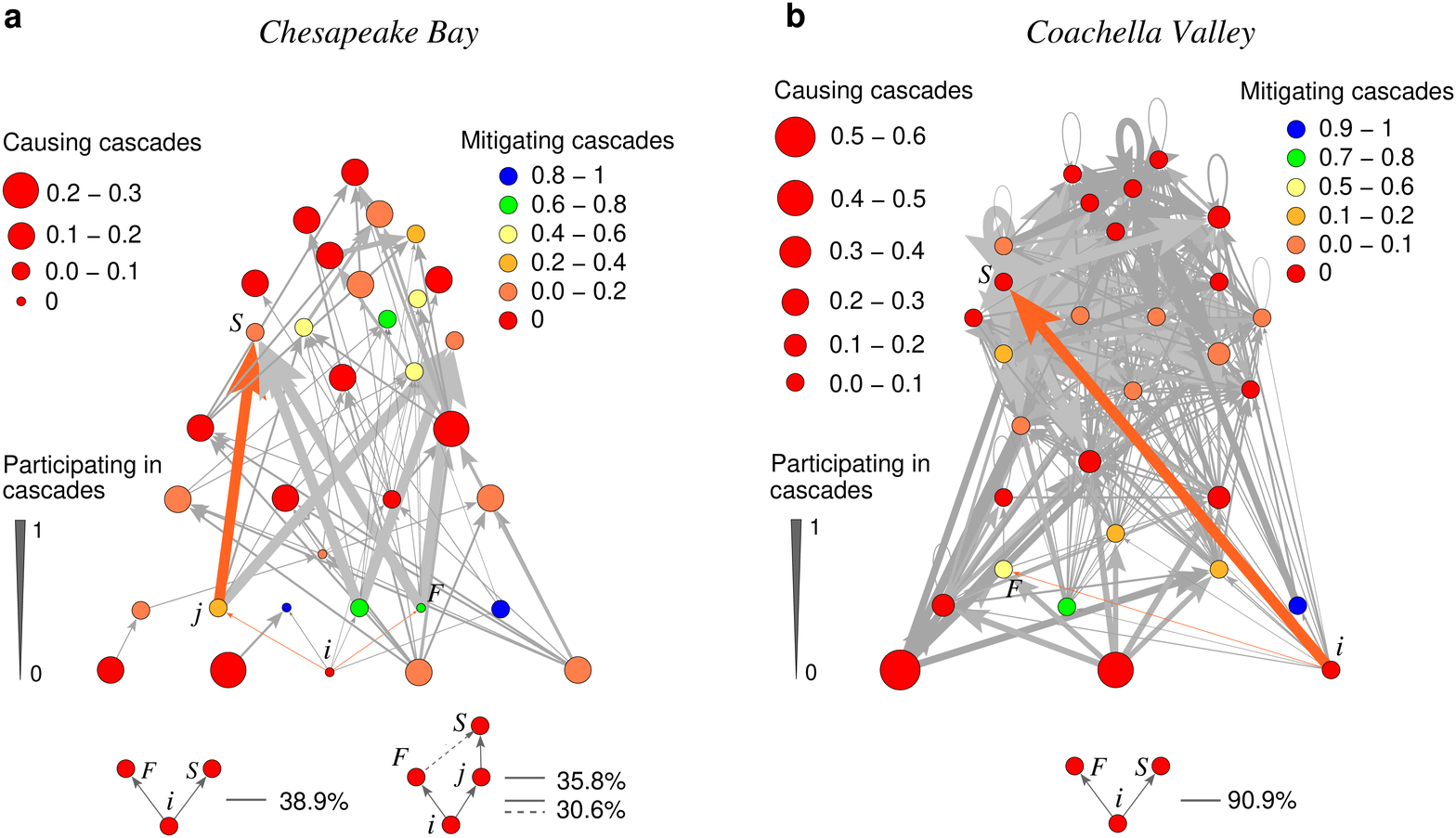}
\caption{ \baselineskip 14pt
\footnotesize{{\bf Extinction cascades and rescue mechanisms predicted for two empirically-observed food webs.} ({\bf a}, {\bf b})  The Chesapeake Bay food web ({\bf a}) and the Coachella Valley food web ({\bf b}), simulated using the consumer-resource model and organized by trophic levels. The size of each node indicates the probability that its removal along with two others would trigger a cascade of two or more secondary extinctions and the color indicates the probability that its removal would reduce the size of one such cascade. The statistics are calculated within the set of all three-species removals that trigger cascades that can be mitigated, and only for forced removals that rescue the largest number of species. The width of each link indicates the probability that the link would disappear as a result of one of these cascades.  In both food webs, a large fraction of rescues are governed by a mechanism observed in the model networks of Figure~\ref{fig2}, in which $F$ 
 and $S$ share a common prey $i$ (({\bf a}, {\bf b}), bottom sets). In $100\%$ of the rescues in which this network structure is present, the removal of $F$ increases $i$, which helps feed $S$. The Chesapeake Bay food web also exhibits a predominately long-range mechanism where the removal of $F$ causes an increase in a prey $i$, which causes an increase in its predator $j$, which in turn feeds $S$ (({\bf a}), bottom-right set). This mechanism is confirmed in $100\%$ of the cases when $S$ is not directly connect to $F$ and in $66.8\%$ of the cases when $S$ feeds on $F$ (dotted line). The percentages in the figure indicate the fractions of forced removals that exhibit the corresponding structures. The orange-coloured links in the main panels correspond to specific examples of these network structures. Complementary statistics are provided in Supplementary Table S1.}
\label{fig4}}
\end{figure}

The Chesapeake Bay and Coachella Valley food webs are explicitly analyzed in Figure~\ref{fig4}.  The former is sparsely connected, has no loops and has a large number of top predators (Fig.~\ref{fig4}a), whereas the latter is densely connected, has loops (including cannibalistic links), and has no top predator (Fig.~\ref{fig4}b).  For both food webs, under the conditions considered in our study, the random assignment of initial populations leads to a single time-independent state, which we perturbed by all three-species primary removals and tested against all single-species forced removals. Figure~\ref{fig4} represents the average over all such independent realizations for which a cascade can be mitigated (835 in Fig.~\ref{fig4}a and 283 in Fig.~\ref{fig4}b), where the probability that a species removal will mitigate or participate in triggering a cascade is coded in the color and size of the nodes, respectively, while the probability that a feeding interaction is eliminated 
 by a cascade is coded in the width of the links. As in other parts of this study, the rescue statistics are drawn from all forced removals that prevent the largest number of extinctions in the given cascade. In both networks, a group of only $2$ low trophic level non-basal species is responsible for rescuing over $96.2\%$ (Chesapeake Bay) and $99.3\%$ (Coachella Valley) of all mitigated cascades (a fraction of these cascades can be equally well mitigated by other forced removals). Note that the network positions of these species are not too different from the basal ones that are among the most likely to cause cascades. The rescues in the Chesapeake Bay food web are frequently determined by a long-range mechanism where the closest interaction is by means of a prey of $S$ that feeds on a prey of $F$, which, counterintuitively, remains frequent even when $S$ also feeds on $F$  (Fig.~\ref{fig4}a). In the Coachella Valley food web, on the other hand, the rescues most frequently 
 involve a mechanism, already identified in the model networks, in which $F$ and $S$ share a common prey species (Fig.~\ref{fig4}b). It is interesting to notice that while our approach can reveal these rescue interactions, they are by no means evident from the network structure alone.

\medskip
\noindent
{\bf Rescuable and non-rescuable species.}
Our results raise the fundamental question of identifying the species that can be rescued by these interventions. For this purpose, we note that all cascading extinctions can be classified into structural extinctions and dynamical extinctions. A structural extinction occurs when a species is left with no directed paths connecting it to basal species in the food web \cite{Stefano}.  A dynamical extinction, on the other hand, is not directly caused by connectivity limitations and is instead determined by the dynamical evolution of the food web. By their own nature of constraining system parameters or variables, the rescue interventions considered in this study cannot prevent structural extinctions. However, they can,
at least in principle, prevent dynamical extinctions. 

Irrespective of being dynamical or structural, cascading extinctions occur when the trajectory describing the evolution of the system falls into the basin of attraction of an attractor for which one or more species have zero population (Methods).
Once found to be in a given basin of attraction, the occurrence of subsequent extinctions is entirely determined, and can only be prevented by rescue interventions such as those considered in this study. A rescue perturbation shifts the state of the system to the basin of an attractor (e.g., a stable fixed point) with a larger number of nonzero-population species, and this is only possible in the case of dynamical extinctions. The size of a basin of attraction, and hence the occurrence of dynamical (but not structural) extinctions, depends on the minimum viable population size \cite{Sh}, which is accounted for by a threshold $s$ in our models (Methods).
This can be rationalized by categorizing the dynamical extinctions into those due to systematic population decrease and those due to oscillations. While the former do not directly depend on $s$, some of the latter extinctions may be absent for smaller $s$. Most importantly, our numerical experiments demonstrate that both types of dynamical extinctions can be mitigated by the rescue perturbations we consider.

For all scenarios considered in our study of model networks, more than 74\% of the cascading extinctions are dynamical, and hence potentially preventable. This fraction is larger for the Lotka-Volterra than for the consumer-resource model. In the latter case we also considered the effect of other parameters, and observed that this fraction is larger for smaller number of primary removals, for larger connectance, and for larger fraction of vertebrate species. All this is consistent with the observed increased availability of basal species and of possible network paths to reach them from other species. In many cases the number of species rescued corresponds to the theoretical maximum (Supplementary Methods), which is generally possible for cascades that only involve dynamical extinctions.
The fraction of cascades mitigated (to any extent) by forced removals tends to exhibit very weak dependence on connectance and tends to decrease with an increase in the size of the primary perturbation. There is also a small dependence on the type of community, where the fraction of mitigated cascades is higher for vertebrates, and the difference matches the corresponding difference in the fraction of purely structural cascades. Very importantly, the fraction of mitigated cascades is observed to consistently increase as the number of species in the network is increased, for both the Lotka-Volterra and the consumer-resource model and for all forms of rescue intervention considered. This provides evidence that the potential benefit of rescue perturbations can in fact be more pronounced for larger food webs. For details on the parameter dependence, see 
Supplementary Methods and Supplementary Figures S5-S8.\\

\noindent
{\large\bf Discussion}

Our proof-of-principle analysis provides a theoretical foundation for the study of extinction cascades in which locally deleterious perturbations can partially or completely compensate for other deleterious perturbations. As a context for the interpretation of these results, it should be noted that wildlife population controls in the form of partial or complete removals, growth suppression and mortality increase of target species have been experimentally applied to both invasive and native species in a number of scenarios in which extinction cascades were not explicitly accounted for.

This has been implemented via hunting, fishing, culling, targeted poisoning and non-lethal removals, and is expected to also benefit from fertility control methods in the future \cite{Ch,Sa,Kirkpatrick1985}. Such interventions may be required to correct unbalances introduced by human activity that gave one species advantage over the others, or to mimic the effect of previously removed natural predators in preventing specific populations from crossing the carrying capacity of the area. In addition, several projects involving the total removal of one or more species have been completed successfully. An important example is the recent removal of feral pigs from Santiago Island, in Ecuador, which were introduced just a few years after Darwin's 1835 visit to the archipelago. This successful eradication was completed in the year 2000 and is impressive both because of the area of the island (over 58,000 ha) and the number of individuals removed (over 18,000 pigs) \cite{cruz2005}. An
 other remarkable example is provided by the Channel Islands, in California, where the introduction of feral pigs drove the population of foxes close to extinction by attracting to the islands a native common predator, the golden eagle. To preserve the foxes, both pigs and eagles were completely removed from Santa Cruz, the largest of the Channel's Islands. The management strategy consisted of non-lethal removal of eagles (e.g., by capturing them with net guns) followed by the lethal removal of the pigs. The order of the removals has been shown to be critical for the preservation of the island fox \cite{collins2009} and was defined based on food-web models of the same nature of those used in our study \cite{Wo}, thus illustrating the usefulness of such models in management decisions. 

By accounting for the cascading consequences of network perturbations, our study addresses an important new aspect involved in these population control applications. We showed that such consequences are often counterintuitive and hence difficult to anticipate from qualitative analysis. An important element in our analysis is the reliability with which the dynamics of the species' populations can be forecast. Another key element is the feasibility of the interventions themselves.
As the above precedents indicate, population suppression and the other interventions considered in our analysis should be interpreted as limited to islands, lakes, parks and other local areas, without involving the large-scale eradication of any species. They may be implemented in concert with economical activities, such as fishing and hunting, but may also be carried out by means of non-lethal growth suppression and relocation. These conclusions are not limited to extinction cascades triggered by an extinction event and can be extended, for example, to mitigate the impact of invasive species.

Taken together, our results suggest that rescue interactions permitting compensatory perturbations are common in food-web systems and that the identification of such interactions can both benefit from accurate food-web models \cite{Wi2000,ca2004,All2008,gross2009} and help constrain such models for stability \cite{may1972,may2009,holland2008,allesina2008,perotti2009}.
These results also provide evidence for the growing understanding that preservation requires more than the absence of active destruction \cite{natedit}, and promise to offer important insights in combination with ongoing projects on proactive management actions, such as assisted migration \cite{hoegh2008}. \\

\noindent
{\large\bf Methods}

\medskip
\noindent
{\bf Target states.}
In the case of the Lotka-Volterra system,
$\dot{X_i}=X_i (b_i +\sum_j a_{ij}X_j)$, $i=1,\cdots, n$,  our approach can be formalized
in terms of the properties of asymptotic stationary states associated with fixed points 
of the dynamics.  The fixed points are given by the
set of equations $X_{i}(b_{i}+\sum_{j}a_{ij}X_{j})=0$, which can be factored 
as 
\begin{eqnarray}
\label{seq1}
b_{i}+\sum_{j}a_{ij}X_{j}&=&0,\\
\mbox{and/or} \;\;\; X_{i}&=&0,
\label{seq2}
\end{eqnarray}
where $i=1,\cdots, n$.
We denote by ${\bf X^*}$ = ($X_{1}^*$, \dots, $X_{n}^*$) the fixed points that correspond to
valid solutions of Eqs.\ (\ref{seq1})-(\ref{seq2}), i.e., solutions for which all populations 
are non-negative.
This set of equations always has at least one solution, namely 
the solution for which only Eq.\ (\ref{seq2}) is satisfied, and all populations are zero. 
In most cases, however, a large number of other valid solutions exist (up to $2^n$ if matrix 
$A=(a_{ij})$ is invertible, as observed for the randomly generated
connection strengths considered in our simulations, and up to infinitely many if 
 $A$ is singular). 
Given a species removal perturbation that triggers a cascade of extinctions, we denote 
by $n^p$ the  
number of nonzero-population species shortly after the perturbation, by $n^c$ the 
number of nonzero-population 
species after the cascade, and by $n^*$ the number of species with nonzero population at 
fixed points 
that correspond to valid solutions consistent with the constraints imposed by the primary
perturbation. Following a perturbation, we refer to such fixed points that in addition satisfy
$n^*>n^c$ as {\it target states}.
The corresponding populations shortly after the perturbation and after the cascade 
are denoted by $X_i^p$ and $X_i^c$, respectively;  hereafter we use $X_i^*$ and $n^*$ specifically to 
denote the populations and number of persistent species at target states. 
Given that the number of nonzero-population
species prior to the perturbation is $n$, it follows that $n^c < n^p < n$ and $n^c<n^*\le n^p$.

\medskip
\noindent
{\bf Rescue interventions.}
Our rescue strategy is based on identifying (and proactively driving the system towards)
a target state, thus preventing the extinction of one or more species. We considered the 
following algorithmic implementations of this concept.

\begin{description} \itemsep -0.3mm
\item[(a)] {\it Forced species removal}. 
Focusing on fixed points that satisfy the constraints imposed by the primary removal,
we first identify all target states with $n^*<n^p$.
We then test one-by-one
the forced removal of each species $i$ for which $X_i^*=0$ at one or more of these states. The forced
removals are implemented shortly after the primary removal, hence before the propagation of 
the cascade. The system is then evolved to test the impact of these removals.
We select the tested removals that lead to the largest reduction in the number of
secondary extinctions. This approach is applicable to cases in which $n^c \le n^p-2$, i.e., 
if the cascade would otherwise consist of two or more extinctions.

\item[(b)] {\it Forced removal of a cascading species}. In this case we test one-by-one the 
forced removal of every species that would be extinct by the cascade, and select the removals
that lead to the largest reduction in the number of secondary extinctions. This is thus similar 
to the approach just described, except that it is now limited to the set of all cascading species. 

\item[(c)] {\it Forced partial removals}. Starting from the target states with the smallest number
of zero-population species (including possibly states with $n^*=n^p$), we test the concurrent
partial removal of all species $i$ for which $X_i^*<X_i^p$ at the given fixed point. 
These partial 
removals consist of reducing the 
population of species $i$ to $X_i^*$ shortly after the primary removal (total removals
are thus included whenever $X_i^*=0$).
The partial removals are tested for all target states  
on a one-by-one basis. 
We select the sets of partial removals that lead to
the largest reduction in the number of secondary extinctions. This approach is also 
applicable to cases in which $n^c = n^p-1$.

\item[(d)] {\it Manipulation of growth and mortality rates}. We test the impact of
growth rate reduction and mortality rate increase by seeking new rates  $b_i^*$
that solve the equation
\begin{eqnarray}
b^*_{i}+\sum_{j}a_{ij}X^{p}_{j}&=&0
\end{eqnarray}
under the constraints that $b^*_i\le b_i$ if $b_i<0$ (mortality rate)
and $0\le b^*_i\le b_i$ if $b_i>0$ (growth rate). 
For all species $i$ that admit such a solution,
the corresponding $b_i$ are
replaced by $b^*_i$ right after the primary perturbation, with the others kept unchanged.
The system is then evolved to determine the impact of this modification
in reducing the number of secondary extinctions.  This modification increases the
likelihood that the asymptotic dynamics of the corresponding species will satisfy 
Eq.\ (\ref{seq1}) or exhibit time-dependent behavior, as opposed to satisfying Eq.\ (\ref{seq2}).
This approach is applicable to cases in which $n^c \le n^p-1$, as it does not
involve the removal of any species.

\end{description}

The rationale underlying these interventions is that by setting the parameters of one
or few species at the values of a fixed point with a reduced number of extinctions,
the system will often
evolve towards that fixed point (or towards an attractor in a common subspace if the 
fixed point is unstable), and secondary extinctions
will be mitigated. Indeed, as shown in this study, the asymptotic 
number $n^*_{new}$ of nonzero-population species 
after any of these compensatory perturbations is often larger than $n^c$.
In general, $n^*_{new}$ is smaller than or equal to the number $n^*$ of nonzero populations 
at the fixed points we target, and the inequality arises 
from the fact that the system may evolve to a different fixed point or other attractor with additional zeros.

\medskip
\noindent
{\bf Stability considerations.}
The 
stability of the fixed points is determined by the eigenvalues of the corresponding
Jacobian matrix 
$J=(J_{ik})$, where $J_{ik} =  X^*_ia_{ik} + \left(b^*_i +\sum_j a_{ij}X^*_j\right)\delta_{ik}$, 
and $b_i^*$ is now used to denote both modified and unmodified rates at the fixed points. 
Note that, at a fixed point, the second term on the r.h.s.\ of the Jacobian matrix is zero for all 
species $i$ for which $X^*_i>0$. 
We thus focus on the 
reduced Jacobian matrix $\tilde{J}=(\tilde{J}_{ik})$ determined by the nonzero-population species, 
\begin{equation}
\tilde{J}_{ik} =  X^*_ia_{ik},
\label{seq3}
\end{equation}
where the indexes $\{i,k\}$ range over all $n^*$ species for which $X^*_i, X^*_k>0$. 
A fixed point ${\bf X^*}$ is linearly stable (unstable) if the real part of the largest eigenvalue 
of $\tilde{J}=\tilde{J}({\bf X^*})$ is negative (positive). Rescue interventions based
on forced removals can be effective even for unstable fixed points because of attractors with 
a reduced number of extinctions that may exist in a common subspace (Supplementary Methods).

\medskip
\noindent
{\bf Effect of minimum population size.}
The reduction of the Jacobian matrix from $J_{n\times n}$ to $\tilde{J}_{n^*\times n^*}$ 
is consistent with our selection of a threshold $s$ for the populations below which they are
assumed to vanish, which is not accounted for by linear stability alone. Biologically, this
represents the irreversibility of an extinction process. Mathematically, this means that
the dynamics of species $i$ is governed by 
$\dot{X_i}=X_i (b_i +\sum_j a_{ij}X_j)$ if $X_{i}\ge s$
for all previous times and by $X_i=0 $ if $X_{i}< s$ 
at any previous time, where in our simulations we take $s$ to be the same for all species.  Therefore, we 
regard $X_i$ to be permanently zero and the Jacobian matrix to be reduced once $X_i<s$. 
The evolution of individual species and the occurrence of extinctions
depend on the threshold value $s$. This simply
indicates that the 
robustness of the system depends on the minimum viable population size.
In the language
of dynamical systems we can say that the basin of attraction associated with
an attractor defined by the extinction of one or more species will generally 
change with increasing $s$.
However,
once an extinction cascade is triggered, the compensatory interventions
are found to be similarly effective for values of $s$ ranging over many 
orders of magnitude.
In all simulations of both dynamical models presented
in the paper, $s$ was set to be $10^{-3}$.

\newpage

\vspace{0.4cm}
\noindent
{\large\bf Acknowledgments} 

\noindent
The authors thank L.\ Zanella for insightful discussions.
This study was supported by
NSF Grant No.\ DMS-0709212,
NOAA Grant No.\ NA09NMF4630406,
and an Alfred P. Sloan Research Fellowship to A.E.M.

\vspace{0.4cm}
\noindent
{\large\bf Additional information}

\vspace{0.2cm}
\noindent
{\bf Supplementary Information} accompanies the paper, and is available at:\\ 
http://www.nature.com/ncomms/journal/v2/n1/abs/ncomms1163.html

\vspace{0.2cm}
\noindent
{\bf Author Contributions:} A.E.M.\ designed and supervised the research. S.S.\ performed the numerical experiments. 
Both authors contributed to the analysis of the results and the preparation of the manuscript.


\vspace{0.2cm}
\noindent
{\bf  Corresponding author:} Correspondence to Adilson E. Motter (motter@northwestern.edu).

\newpage 



\renewcommand\figurename{{\bf Figure}}




\end{document}